# Comments on Controllable Three-Dimensional Brownian Motors

Elias P. Gyftopoulos
Ford Professor Emeritus
Departments of Nuclear and Mechanical Engineering
Massachusetts Institute of Technology
77 Massachusetts Avenue
Cambridge, Massachusetts  02139 USA

Upon reviewing the physical review letter which describes the processes involved in "demonstration of a controllable three-dimensional Brownian motor in symmetric potentials" we conclude that such processes are not compatible with what Einstein and many other physicists and engineers define as Brownian motors or Brownian movements.

05.40.-a, 82.70.Dd

I studied the letter authored by Sjölund et al [1], and wish them every success in their efforts to explore and analyze whatever practical opportunities exist in the emerging fields of nanotechnology and physiology.  However, I must bring to their attention that Brownian motors or Brownian movements are not involved in the problems they are trying to explore and exploit for the following reasons.

Brownian movement is observed in systems consisting of a liquid solvent and a colloid in a neutral stable equilibrium state, and what appears as Brownian movement is the result rather than the cause of the movements that are observed over long periods of time.  A complete discussion of this statement is given in [2, 3]. For the purpose of my comments, I will summarize only a few key concepts and results.

A system having an amount of energy denoted by $E$, r different constituents with the amounts denoted by $\boldsymbol{n} = \{n_1, n_2, ..., n_r\}$, and a fixed volume denoted by $V$, can be in one of an infinite number of states. However, *the second law of thermodynamics* [4] *asserts that one and only one of these states is a globally stable equilibrium state*. It follows that any property of a system in a stable equilibrium state must be a function only of $E$, $\boldsymbol{n}$, and $V$. In particular, the entropy of a stable equilibrium state must be of the form $S(E, \boldsymbol{n}, V)$ and this form is called the *fundamental relation*. Among many practical applications, the fundamental relation is used for the definitions of *temperature*, *pressure*, and *total potentials*, properties that are valid only for stable equilibrium states. In particular, for a system that has volume as the only parameter, the definitions of the properties just cited are given by the relations

$$\text{Temperature:} \quad 1/T = (\partial S/\partial E)_{\boldsymbol{n}, V} \quad (1)$$

$$\text{Total potentials:} \quad \mu_i = (\partial E/\partial n_i)_{S, \boldsymbol{n}} = -T(\partial S/\partial n_i)_{E, \boldsymbol{n}, V} \quad \text{for } i = 1, 2, ..., r: \quad (2)$$

$$\text{Pressure:} \quad p = T(\partial S/\partial V)_{E, \boldsymbol{n}} \quad (3)$$

It is noteworthy that $T$ and $\mu_i$'s are interpreted as measures of escaping tendencies in the following sense. Given two systems A and B in stable equilibrium states, energy and entropy flow from A to B if and only if $1/T_A < 1/T_B$ for both positive and negative $T_A$, and constituent i flows from A to B for $T_A = T_B$ if and only if $(\mu_A)_i > (\mu_B)_i$. Pressure is interpreted as measure of a capturing tendency for volume.

The fundamental relation depends on the volume of a system but not on the shape of the volume. For example, two systems containing identical types and amounts of constituents, one being a cube of volume $V_c$ and the other an orthogonal prism of volume $V_p = V_c$, have identical fundamental relations. It follows that

$$(dS)_{E, \mathbf{n}, V} = S_c - S_p = 0 \qquad (4)$$

To be sure the preceding equality is valid for any pair of equal volumes regardless of the size and the shape of each volume.

Systems that satisfy Eq. 4 for any pair of equal volumes have been discussed by Hatsopoulos and Keenan [5], and are said to be in *neutral stable equilibrium*, the neutrality referring to the constancies of the energy, amounts and types of constituents, and the size of the volume regardless of its geometric shape. But we must recognize that changes of shape play a decisive role in determining how the parameters affect the behavior of the system and, as we will see, in the theoretical understanding of Brownian movement.

For example, for a cubical shape of side equal to a, the volume $V_c = a^3$ and only the parameter "a" determines the details that enter in the evaluation of the value of the fundamental relation. On the other hand, if the shape is an orthogonal prism with sides b, c, and d, and volume $V_p = bcd = a^3 = V_c$ (Figure 1), then there are three parameters that enter in the evaluation of the fundamental relation. Even though this change of parameters does not affect the value of the fundamental relation, it affects the configuration of the constituents which in the cube is determined only by "a", whereas in the prism it is determined by "b", "c", and "d". We will see later that the change of configuration is continuous in time, and is evidenced as Brownian movement. Another example of changes of shapes is represented by the white background and the black caricatures in Figure 2.

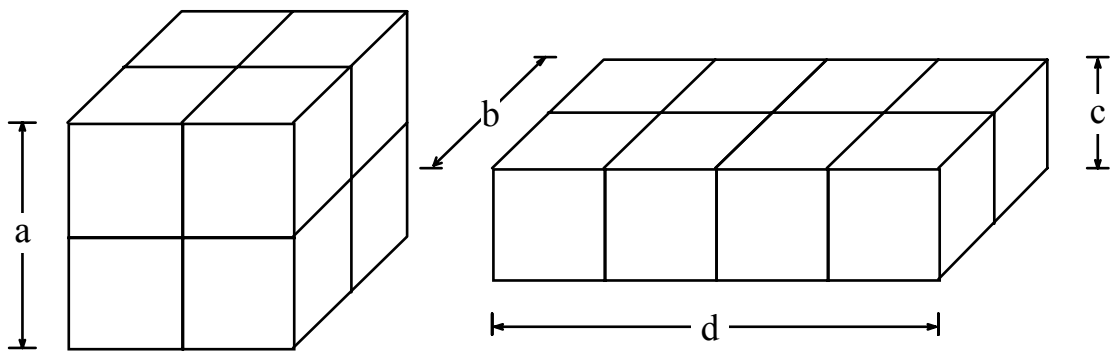

**Figure 1: A simple example of shape change without volume change.**

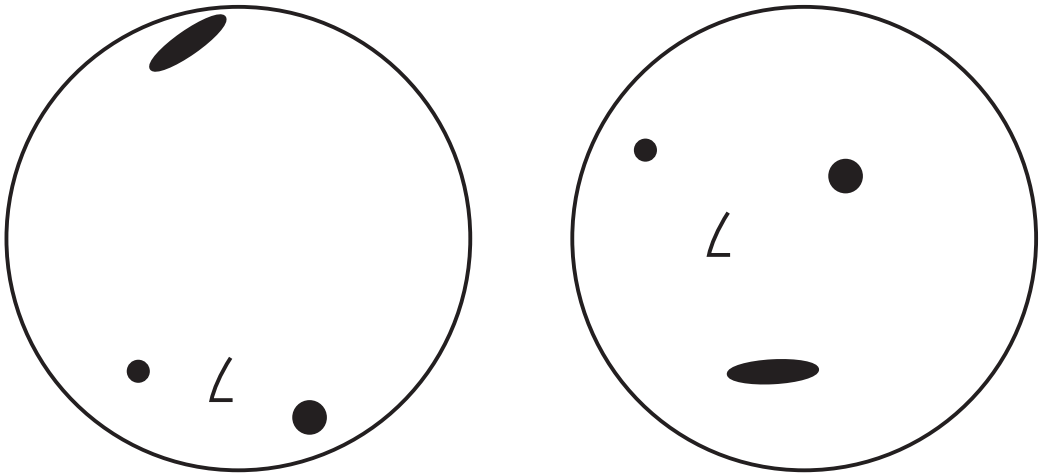

**Figure 2: A realistic example of shape changes of a solvent (white) and a colloid (black spots) without changes of the volumes of either the solvent or the colloid.**

Systems in which Brownian movement is observed are in a neutral thermodynamic equilibrium state and consist of two phases: (i) a *liquid solvent* capable of dissolving and/or dispersing one or more other phases; and (ii) a colloid composed of particles much larger than atoms or ordinary molecules but much too small to be visible to an unaided eye, dispersed but not dissolvable in (i).

For our purposes, we consider an isolated composite system consisting of a liquid solvent A, and a dispersed but not dissolvable colloid B in a neutral stable equilibrium state that has energy $E = E_A + E_B$, amounts of constituents $\bm{n} = \bm{n}_A + \bm{n}_B$, volume $V = V_A + V_B$, temperature T, pressure p, and $E_i$, $\bm{n}_i$ and $V_i$ have fixed values for i either A or B. Persistent experimental observations indicate that in such a system the constituents of a colloid and, consequently, also of the liquid solvent appear to aimlessly, and for all practical purposes endlessly be moving around over observed time periods of many years. This is what Einstein and many other scientists call Brownian movement, and there is no doubt whatsoever about the existence of the phenomenon.

In our work, we have proven that in a system in a stable equilibrium state no particle of any constituent is moving – each particle has zero velocity [6, 7]. Said differently, we have shown that Maxwell's demon cannot accomplish his task because there are no fast and slow atoms or molecules to be sorted out. So one might jump to the conclusion that Brownian movement contradicts this result. However, the contradiction is illusory for the following reason.

In thermodynamics, we find that each total potential is interpreted as an escaping tendency [4]. Moreover, we prove that if a constituent is not present in a system then its total potential is equal to minus infinity [8]. So for the solvent and the colloid the following relations apply for each constituent i of the solvent, and j of the colloid:

$$(\mu_{\text{solvent}})_i > (\mu_{\text{colloid}})_i = -\infty \tag{5}$$

$$(\mu_{\text{colloid}})_j > (\mu_{\text{solvent}})_j = -\infty \tag{6}$$

Inequalities 5 and 6 show that systems in which Brownian movement is observed are in partial mutual stable equilibrium, that is, they satisfy the conditions of temperature and pressure equalities but not the conditions of total potential equalities. As a result, the constituents of both the solvent and the colloid exert infinitely large "driving forces" (total potential differences) on the interface between the two phases, and try to interpenetrate each other as they would have done if the colloid were soluble by the solvent. However, such interpenetration is impossible, and the only effect is a continuous in time modification of the pliable shape of the interface between the two phases, the battle goes on forever and appears to any observer as Brownian movement. Said differently, it is not phase motions that cause the observed movements but infinitely large differences in total potentials that change the shape of the interface and appear as motions.

In the quantum mechanical analysis of the problem [2] we provide explicit results of the effects of continuous interfacial shape changes. We consider the same system defined for the thermodynamic analysis of Brownian movement but describe the quantum probabilities by a density operator that involves no statistics of statistical quantum mechanics. Said differently, the density operator $\rho$ we use is represented by a homogeneous ensemble each member of which is characterized by the same $\rho$ [2, 9]. The constituents of both the liquid solvent and the colloid worm aimlessly and endlessly their ways within each other because of the infinitely large differences between total potentials (Eqs. 5, 6), while each phase is passing through a neutral stable equilibrium state, that is, the energy, amounts of constituents, and volume of each phase is fixed but the parameters are changing in a very complex and sinuous manner. As a result, the energy eigenprojectors and eigenvalues of both the liquid solvent and the colloid, and the corresponding stable equilibrium state density operators change continuously in time so as to accommodate the continuously changing shapes – parameters – of the volumes of the two systems of the composite of the liquid solvent and the dispersed colloid. These continuous changes in time are impossible to evaluate because both of the lack of knowledge of the precise change of the shapes of the volumes, and the difficulty inherent in calculating energy eigenprojectors and eigenvalues in cases of complicated shapes of even very simple systems such as one particle in an odd looking, one dimensional potential well. The laws of physics, however, have no difficulty in continuously in time responding to the changing shapes of the liquid solvent and the colloid and determining the nonstatistical density operators for each pair of shapes at each instant in time.

An illustration of the quantum mechanical effects of the changes of the shapes of the volumes of a composite system consisting of a solvent and a colloid in neutral stable equilibrium states has been made by Çubukçu [10]. He considers the Hamiltonian operators $H_g$ for g = s or c, where s is the solvent and c the colloid, and proceeds with the following calculations:

Hamiltonian operator of the two systems

$$H = H_s \otimes I_c + I_s \otimes H_c \qquad (7)$$

where I is the identity operator

Energy eigenfunctions and eigenvalues

$$H_s\psi_i = e_i\psi_i \qquad H_c\varphi_j = \varepsilon_j\varphi_j \tag{8}$$

$$H(\psi_i \otimes \varphi_j) = (e_i + \varepsilon_j)(\psi_i \otimes \varphi_j) \tag{9}$$

Density operator of both phases

$$\rho(t) = \rho_s(t) \otimes \rho_c(t) \tag{10}$$

Eigenvalues of density operator

$$\rho(t)(\psi_i \otimes \varphi_j) = p_{ij}(\psi_i \otimes \varphi_j) \tag{11}$$

Relation between density operator eigenvalues and energy eigenvalues

$$\ln(p_{ij}/p_{kl}) = [(e_k - e_i) + (\varepsilon_l - \varepsilon_j)]/kT \tag{12}$$

for all pairs of indices $\{i, j\}$ and $\{k, l\}$, where $T$ is the constant temperature of the solvent and the colloid.

From Eqs. 7 to 12 we see that as the eigenvalues and eigenfunctions of the solvent and the colloid change, the constituents of each of the two phases are continuously reallocated to the evolving energy eigenstates, and the reallocation appears as Brownian movement.

Said differently, the reallocations are the cause of the motions, and not the motions the cause of the reallocations.

Despite claims to the contrary [1, 11], physiological phenomena, such as *E. coli*, are not related to Brownian movements or Brownian motors. In contrast to systems in which Brownian movement is observed and which consist of a liquid solvent and a colloid, each maintaining its identity for all practical purposes for ever, *E. coli* have a totally different biography. Relevant statements that one finds in textbooks on molecular biology of the gene [12] and biochemistry [13] are as follows: "… the bacteria *E. coli* will grow in an aqueous solution containing just glucose and several inorganic ions" … "There is a lower limit, however, to the time necessary for a cell generation; no matter how favourable the growth conditions, an *E. coli* is unable to divide more than once every 20 minutes. … The average *E. coli* cell is rod shaped. … It grows by increasing in length followed by a fission process that generates two cells of equal length." In addition, *E. coli* is self propelled not as a result of infinite differences between total potentials such as exist between a liquid and a solvent and a colloid but because of flagella.

In view of all these facts, one must conclude that the time dependent changes of *E. coli* studied by molecular biologists, and biochemists do not represent Brownian motions. They are the result of chemical reactions between *E. coli* and the nutrients supplied by the solvent.

In closing this letter, I wish to express my hope that my remarks, and the correct use of thermodynamics and quantum thermodynamics without statistical probabilities will be helpful to the authors of references [1] and [11], as well as researchers of other phenomena both at the nanoscale level and larger sizes.

Electronic address: dboomer@mit.edu